\documentclass[final,5p]{elsarticle}
\usepackage{hyperref,booktabs,subfigure,widetext,caption,multicol,natbib}

\usepackage{mymacros}
\usepackage{amssymb, amsmath}	%% Added by YI
\input{colordvi.tex}	%% Added by YI

\journal{Journal of \LaTeX\ Templates}

\biboptions{numbers,sort&compress}
\begin{document}

\begin{frontmatter}

\title{Cherenkov Telescope Array extragalactic survey discovery potential and the impact of axion-like particles and secondary gamma rays.}

\author[rvt]{Andrea De Franco\corref{cor1}\fnref{fn1}}
\ead{andrea.defranco@physics.ox.ac.uk}
\author[focal]{Yoshiyuki Inoue}
\author[els]{Miguel A. S\'anchez-Conde}
\author[rvt]{Garret Cotter}
\cortext[cor1]{Corresponding author}
\fntext[fn1]{Supported by the 2012-FP7-ITN, nr 317446, INFIERI EU programme.}
\address[rvt]{University of Oxford, department of Astrophysics, Parks Road,Oxford,UK}
\address[focal]{Institute of Space and Astronautical Science JAXA, 3-1-1 Yoshinodai, Chuo-ku, Sagamihara, Kanagawa 252-5210, Japan}
\address[els]{Oskar Klein Centre for Cosmoparticle Physics, Department of Physics, Stockholm University, SE-10691 Stockholm, Sweden}
\address[els]{Instituto de F\'{\i}sica Te\'orica UAM/CSIC, Universidad Aut\'onoma de Madrid, E-28049 Madrid, Spain}
\address[els]{Departamento de F\'isica Te\'orica, M-15, Universidad Aut\'onoma de Madrid, E-28049 Madrid, Spain}

\begin{abstract}
The Cherenkov Telescope Array (CTA) is about to enter construction phase and one of its main key science projects is to perform an unbiased survey in search of extragalactic sources. We make use of both the latest blazar gamma--ray luminosity function and spectral energy distribution to derive the expected number of detectable sources for both the planned Northern and Southern arrays of the CTA observatory. We find that a shallow, wide survey of about 0.5 hour per field of view would lead to the highest number of blazar detections. Furthermore, we investigate the effect of axion-like particles and secondary gamma rays from propagating cosmic rays on the source count distribution, since these processes predict different spectral shape from standard extragalactic background light attenuation. We can generally expect more distant objects in the secondary gamma-ray scenario, while axion-like particles do not significantly alter the expected distribution. Yet, we find that, these results strongly depend on the assumed magnetic field strength during the propagation. We also provide source count predictions for the High Altitude Water Cherenkov observatory (HAWC), the Large High Altitude Air Shower Observatory (LHAASO) and a novel proposal of a hybrid detector.
\end{abstract}


\end{frontmatter}

\section{Introduction}\label{sec:Introduction}
Current generation instruments detected a few hundred objects in the very high energy (VHE) gamma--ray band above 50 GeV during the past decade, a good fraction of which also come with a distance measurement from multi--wavelength campaigns and optical observations~\footnote{\url{http://tevcat.uchicago.edu/}}~\cite{FermiCatalog3,FermiCatalog2,FermiCatalog1,VERITASCatalog,HESSCatalog}. The Cherenkov telescope array (CTA) is expected to come online within the next few years and outperform current imaging atmospheric Cherenkov telescopes (IACTs). Its improved flux sensitivity and larger field of view (FoV) will enable the detection of many new sources and will allow for population oriented studies at VHE~\cite{Dubus2013}. An up-to-date estimate of detectable sources is needed in order to devise the best CTA survey strategies.

Blazars, a class of active galactic nuclei (AGNs), are the dominant population in the VHE gamma--ray sky. Currently, $\sim200$ VHE blazars have been reported~\cite{Fermi2FHL} out to redshift $z\sim1$~\cite{Tanaka2013,Ahnen2015,Abeysekara2015}. The number of VHE blazars is expected to dramatically increase in an improved IACT sensitivity. Indeed, it would be possible to perform a statistical study of VHE blazars in the CTA era (see, e.g., Inoue \& Tanaka 2016 using current IACT blazar samples~\cite{Inoue2016}), which would provide the key to understand AGN populations in the VHE end and high-energy phenomena in the vicinity of supermassive black holes. 

The expected potential of this upcoming VHE blazar CTA survey has been studied in the literature (Inoue, Totani, \& Mori 2010~\cite{Inoue2010}, Dubus et al. 2013~\cite{Dubus2013}, Inoue, Kalashev, \& Kusenko 2014~\cite{Yoshi2ndGamma}, and more). Very recently, Ajello et al.~\cite{Ajello2015} reported an improved model of the spectral energy distributions (SEDs) and evolution (gamma-ray luminosity function, GLF) of blazars based on the latest catalog from the Large Area Telescope~\cite{LATFermi} on board the NASA {\it Fermi} gamma--ray satellite. Utilizing these latest models, in this work we will show the expected source count distribution and redshift distribution at the energy bands covered by the CTA sensitivity. We will also consider the cases for the high altitude water Cherenkov (HAWC)~\cite{HAWCDesign} and the large high altitude air shower observatory (LHAASO)~\cite{LHASSODesign,LHASSODesign_2,LHASSOProspect}. We include consideration for a recent proposal of a hybrid detector composed of a carpet of resistive plate chambers on top of an array of water Cherenkov detectors dubbed Hybrid in this work~\cite{Hybrid}.

VHE gamma rays propagating in intergalactic space are known to be attenuated by the extragalactic background light (EBL)~\cite{EBL_1,EBL_2,EBL_3,EBL_4}. EBL attenuation would affect the expected number of extragalactic source detections in a future survey. Recent studies have detected attenuation of gamma rays on EBL~\cite{EBL_VHE_1,EBL_VHE_2,EBL_VHE_3}, using a dataset dominated mostly by optical depth of the order of unity. However, distant VHE blazars appear to have harder intrinsic spectra than simple gamma-ray emission models~\cite{Horns2012}, as well as a redshift dependence of the observed spectral index that is different from what was expected~\cite{Blazar_z_1,Blazar_z_2,Rubtsov2014,Sanchez2013A&A}, although a large uncertainty remains in the measured redshifts and spectral indices~\cite{ Costamante2013,Dominguez_Ajello2015}. 

To explain such intrinsically hard spectra, several scenarios have been proposed in the literature such as secondary cascade components generated by very high energy cosmic rays~\cite{Acero2013,EsseyKusenko2010,Essey2010}, emission from stochastically accelerated particles in the jet~\cite{Lefa2011}, axion-like particles (ALPs)~\cite{Doro2013,Abramowski2013,deAngelis2011,deAngelis07,Mirizzi07,MiguelALP,Meyer2013PhRvD} and Lorentz invariance violation~\cite{Doro2013}. Except for the stochastic acceleration scenarios, that also suffer from EBL attenuation, the other mentioned processes would affect gamma--ray propagation in intergalactic space, potentially hardening the observed blazar spectrum as compared to the pure gamma-ray absorption on standard EBL scenario. This would be particularly true for sources located at high redshifts. In this paper, we also study implications of ALPs and secondary gamma rays on the source count distribution and whether the number of detected sources with an extragalactic survey with CTA could provide enough statistical significance of one of this effects at work.

This paper is organized as follows. In Section~\ref{sec:Method}, we describe the model used to compute the cumulative source count distribution under the above mentioned scenarios. Discussion of the results is presented in section~\ref{sec:Results}. Conclusions are given in section~\ref{sec:Conclusions}. Throughout the paper we consider the cosmological parameters: H$_0$ = 67 km s$^{-1}$ Mpc$^{-1}$, $\Omega_M$ = 1 - $\Omega_{\Lambda}$ = 0.3.

\begin{figure*}[!htb]
  \centering
\includegraphics[width=0.95\textwidth]{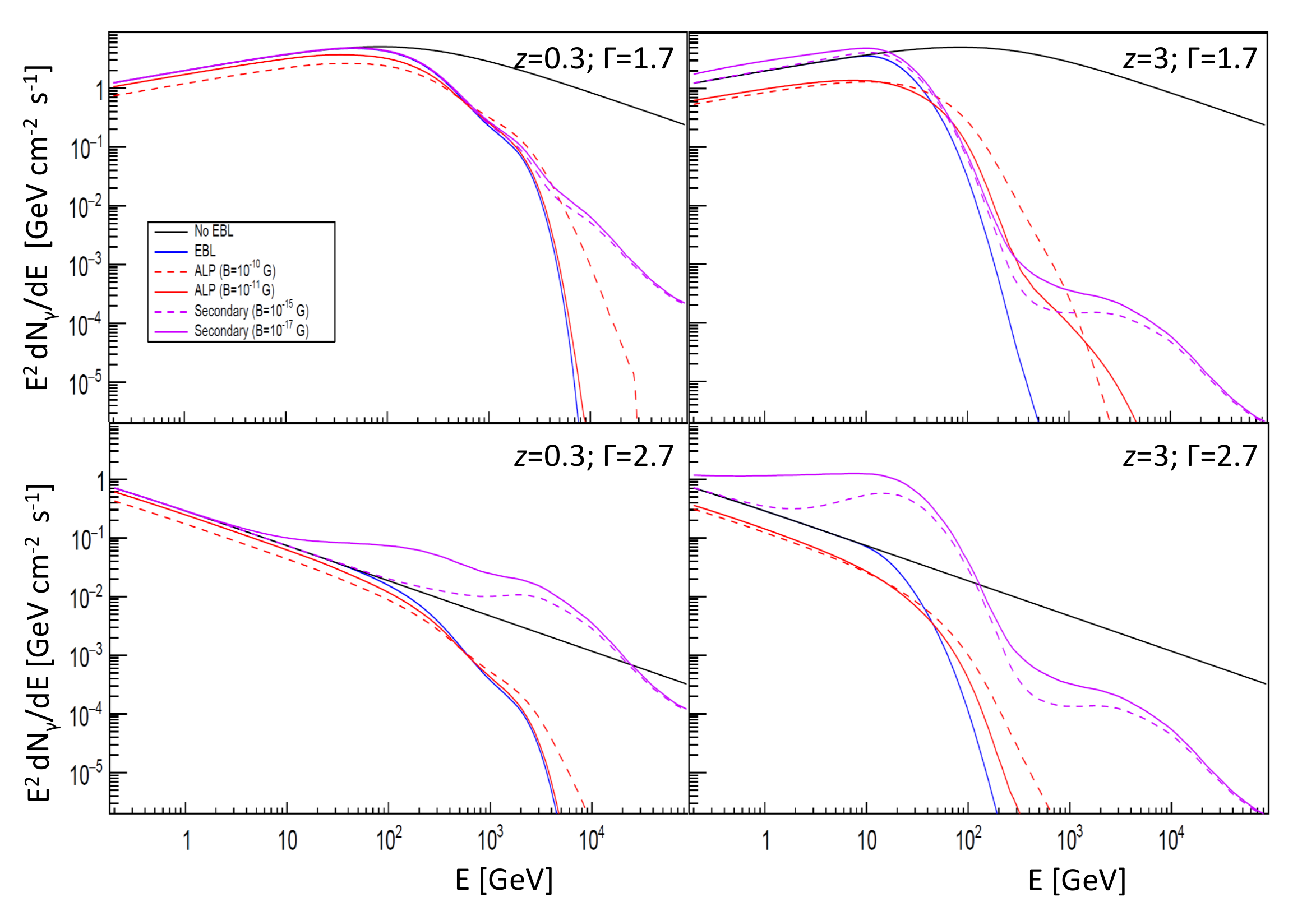} 
\caption{Averaged blazar SED with different photon index $\Gamma$ and redshift z as observed at Earth for four different scenarios: intrinsic spectrum \textit{(black solid line)}; EBL absorption \textit{(blue solid)}; EBL+ALPs \textit{(red lines)}; EBL + secondary gamma rays \textit{(magenta lines)}. In the last two scenarios, dashed and solid curves refer to two different values of the IGMF strength, see legend. Normalization is set by the primary gamma--ray flux at 0.1 GeV, and it is equal to 1 GeV cm$^{-2}$ s$^{-1}$.}
\label{figSED}
\end{figure*}

\begin{figure*}[!htb]
  \centering
\includegraphics[width=0.95\textwidth]{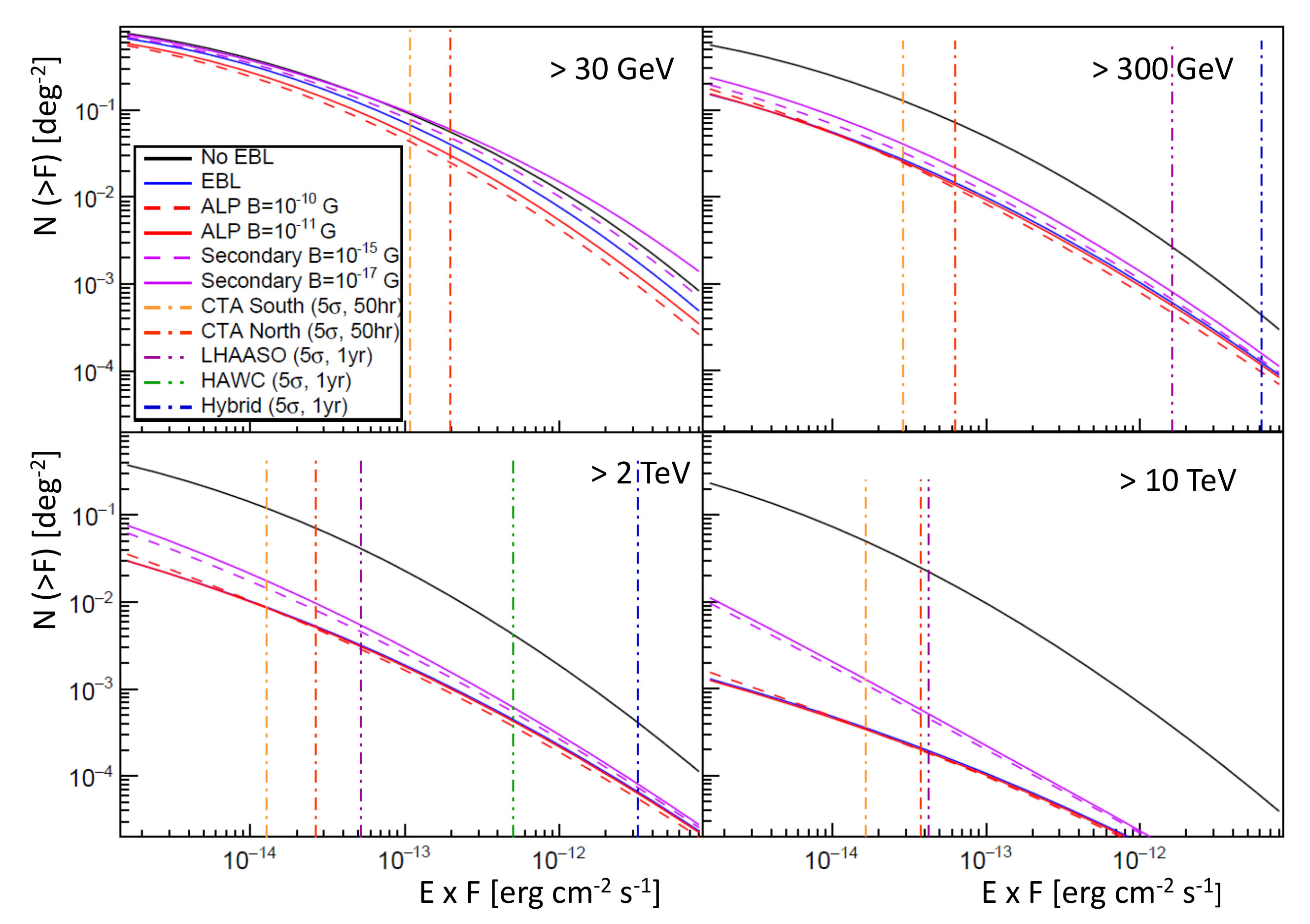} 
\caption{Cumulative source count distribution in different energy bands for the different scenarios discussed in the text: intrinsic spectrum \textit{(black solid line)}; EBL absorption \textit{(blue solid)}; EBL+ALPs \textit{(red lines)}; EBL + secondary gamma rays \textit{(magenta lines)}. In the last two scenarios, dashed and solid curves refer to two different values of the IGMF strength, see legend. In the energy bands in which information is available, we also show the integral flux sensitivity of CTA-South \textit{(vertical, orange dashed-dotted)} and North (5$\sigma$, 50hr observation per FoV) \textit{(vertical, red dashed-dotted)}; Hybrid (5$\sigma$, 1 year) \textit{(vertical, blue dashed-dotted)}; LHAASO (5$\sigma$, 1 year) \textit{(vertical, purple dashed-double dotted)}; HAWC (5$\sigma$, 1 year) \textit{(vertical, green dashed-triple dotted)}.}
\label{figCumulativeCount}
\end{figure*}

\section{Method}\label{sec:Method}
Cumulative source counts above a certain flux limit $F$ is given as 
\begin{equation}
\label{eqSourceCount}
N(>F) = 
\int_{z_{\rm min}}^{z_{\rm max}} \int_{\Gamma_{\rm min}}^{\Gamma_{\rm max}}\int_{L_{min}(F,z,\Gamma)}^{L_{\rm max}} \Phi(L_{\gamma},z,\Gamma) \frac{dV}{dzd\Omega}dL d\Gamma dz,
\end{equation}
where $z$ is the redshift of the source, $\Gamma$ is a photon index (see eq.~\ref{eqEb}), $L_{\gamma}$ is the intrinsic source gamma--ray luminosity in the energy band $0.1-100 \rm GeV$. $\Phi$ is the blazar GLF and $V$ is the comoving volume~\cite{CosmographyReview}. For the GLF, we use the luminosity dependent density evolution (LDDE) model in~\cite{Ajello2015}. We set $z_{\rm min}=0.001$, $z_{\rm max}=4$, $\Gamma_{\rm min}=1$, $\Gamma_{\rm max}=3.5$, $L_{\rm max}=10^{52} \ {\rm erg\ s^{-1}}$ as in~\cite{Ajello2015}. $L_{min}$ is the intrinsic source luminosity corresponding to the flux detected at Earth. We constrain $L_{min}$ to be greater than $10^{42} \ {\rm erg\ s^{-1}}$ following~\cite{Ajello2015}. 
To relate $L_{min}$ to a certain flux limit, we make use of an average blazar SED template presented in \cite{Ajello2015}:

\begin{equation} \label{eqSED}
\frac{dN_\gamma}{dE}(E,\Gamma,z)\propto \left[\left(\frac{E}{E_b}\right)^{1.7}+\left(\frac{E}{E_b}\right)^{2.6}\right]^{-1} e^{-\tau(E,z)},
\end{equation}
where the dependency from measured $\Gamma$ derives from:
\begin{equation} \label{eqEb}
\log E_b ({\rm GeV}) = 9.25 - 4.11\Gamma
\end{equation}
For the opacity coefficient $\tau$ due to absorption of gamma-ray photons on the EBL, we adopt the model of~\cite{YoshiEBL} as our reference one. We verified that there is no significant disagreement in results computed with other EBL models~\cite{DominguezEBL,FinkeEBL,SteckerEBL}. Below, we explore the effect of two different scenarios on the cumulative source count distribution.

\subsection{Axion-like particles}
ALPs are pseudo-scalar bosons similar in properties to standard QCD axions~\cite{CPViolation,PecceiQuinn} but with a coupling constant which is independent of their mass. They are predicted by several extensions of the Standard Model and can constitute all or part of the dark matter density (see, e.g., ~\cite{Axion} for a recent review). Photons couple with ALPs in the presence of external magnetic fields, therefore should such kind of particles exist, VHE gamma rays would oscillate back and forth to ALP in astrophysical magnetic fields during their propagation. 

ALPs would not interact with the EBL, possibly resulting in a less opaque universe to VHE gamma rays. As a result, CTA could observe spectral hardening~\cite{AxionCTA} of distant sources due to this effect.

In Ref.~\cite{MiguelALP} the authors explore different intergalactic magnetic field (IGMF) strength values and conclude that the effect is evident for an IGMF strength of $B=10^{-10}$~G and $10^{-11}$~G, uniform in 1 Mpc domains but whose orientation randomly varies from one domain to another. The probability of photon/ALP conversion increases with the component of the magnetic field along the polarization vector of the photon~\cite{Hooper2007PhRvL}, but the overall effect on the observed spectra also depends on the distance travelled and the precise structure of the magnetic field (direction, strength, size of coherent domains). For example, in cases where the photon/ALP conversion probability is very high, ALPs would not travel long distances before oscillating back to photons, thus not efficiently preventing absorption on EBL. On the other hand, if the mixing effect is not efficient enough, very few photons can oscillate to ALPs, leading to a negligible overall effect on the observed spectra. For simplicity, we adopt uniform IGMF strengths of $B=10^{-10}~{\rm G}$ and $10^{-11}~{\rm G}$ for ALPs in this paper, following the values adopted in Ref.~\cite{MiguelALP}. We note that these values are allowed by current observational constraints on the IGMF strength~\cite{NERONOV}, although they lie close to the existing upper limits coming from the cosmic microwave background (CMB). The spectral distortions induced by photon/ALP mixing on top of the EBL-absorbed source spectrum are computed following~\cite{MiguelALP}. The general effect is a hardening of the spectrum for sources at \textit{z} $\gtrsim0.2$, yet as said this result largely depends on the properties of the IGMF, to a large extent unconstrained at present~\cite{IGMF_1,IGMF_2,IGMF_3,IGMF_4,IGMF_5,IGMF_6,IGMF_7,Pshirkov2016PhRvL}. 

It is important to emphasize that the ALP scenario presented in~\cite{MiguelALP} for the intergalactic case only represents an average case over a large number of realizations of the IGMF configuration (strength and orientation). It has been shown that this average typically leads to a less opaque universe to gamma rays as due to ALPs; however the actual range of possibilities is expected to be much larger and, indeed, may lead to a more opaque universe~\cite{Mirizzi09}. We also note that ALPs/photons oscillation in our Galaxy is not taken into account in our paper, though it could also contribute to alter the spectra significantly~\cite{ALPGalactic1,ALPGalactic2,ALPGalactic3,ALPGalactic4}.

\subsection{Secondary gamma rays}
The origin of high and ultra-high-energy cosmic rays (UHECR) has not yet been uniquely identified, and AGN could contribute to it up to the EeV energies or even higher~\cite{AGNUHECR} (but see also~\cite{TanakaInoue2016b}). Energetic protons propagating in the intergalactic space interact with the intergalactic photon fields via photo-pion production ($p\gamma\rightarrow p\pi^0 /  n\pi^{+}$)  and electron/positron pair production ($p\gamma\rightarrow pe^{-}e^{+}$). Both channels lead to cascades of particles and generate secondary gamma rays, which would be detected along the line of sight of the blazar~\cite{EsseyKusenko2010, Essey2010, Essey2011, Razzaque2012, Murase2012, Aharonian2013, Takami2013, ZhengKang2013}, provided that the IGMF is less than $\sim1~{\rm nG}$~\cite{Essey2011,Murase2012}. 

Protons with energies around $\sim$1~EeV, i.e. below the Greisen--Zatsepin--Kuzmin (GZK)\cite{GZK1,GZK2} cutoff, would not be absorbed by the CMB and could then propagate over cosmological distances before interacting with the EBL photons in the infrared to ultraviolet wavelength. Therefore, secondary gamma rays generated by those protons would be emitted closer to the observer than the primary gamma--ray photons (i.e. the original cosmic--ray acceleration site) and thus would suffer less absorption on the EBL. The flux of secondary gamma rays would add to the source primary flux, in this way enhancing the number of detectable sources for a certain flux sensitivity~\cite{Yoshi2ndGamma}. 

The secondary gamma--ray flux from proton cascade in the intergalactic medium is calculated with the code described in~\cite{OlegCode1,OlegCode2}, where we assume $B=10^{-15}$~G and $10^{-17}$~G for which the secondary photons are not bent significantly and are still detected on the line of sight of the source, and the effect is not too efficient at low energies, so to match fit of blazars spectra. The secondary gamma-ray spectrum shape does not strongly depend on the bolometric proton luminosity of the source, nor on the proton injection spectrum~\cite{Essey2011,EsseyKusenko2010}. However the gamma--ray flux is proportional to the source bolometric luminosity, $L_{p,bol}$. In Ref.~\cite{PBol}, the authors estimated $L_{p,bol}$ to be of the same order, or slightly higher, than $L_{\gamma,bol}$ for BL Lac objects, while being one or two orders of magnitude greater than  $L_{\gamma,bol}$ for FSRQs. The argument assumes a jet populated by one proton per electron, i.e. a negligible content of positron-electron in the jet.  Although a scenario with pairs only in the jet is ruled out by X-ray observation of FSRQs~\cite{Sikora2000}, a certain amount of pairs component in the jet is not yet ruled out~\cite{ Sikora2000,Sikora2005,Bottcher2013,Ghisellini2012}.

Similar analysis was performed in a previous work, that instead used SED template and GLF supported by the EGRET data~\cite{Yoshi2ndGamma}. We make the same assumptions of $L_{p,bol}=L_{\gamma,bol}$ and a proton injection spectrum
\begin{equation} 
\label{eqProtoSED}
j_P(E) \propto E^{-2} \exp(-E/E_{p,max}) \exp(-E_{p,min}/E)
\end{equation}
with $E_{p,max}$ = 1 EeV, $E_{p,min}$ = 0.1 EeV. The lower cutoff may exist due to capture of low energy protons by the local magnetic fields in the source. It does not influence the results, but it is introduced to satisfy source power requirements~\cite{OlegCode1,EminPower} and it can be interpreted as a capture process of low energy protons in the source magnetic field. 

In Fig.~\ref{figSED} we show how EBL absorption, gamma--ray oscillation to ALPs and secondary gamma rays modify the observed spectrum of a source with photon indices of $\Gamma=1.7$ and 2.7 located at redshifts $z=0.3$ and 3.

\begin{figure*}[h!]
  \centering
\includegraphics[width=0.95\textwidth]{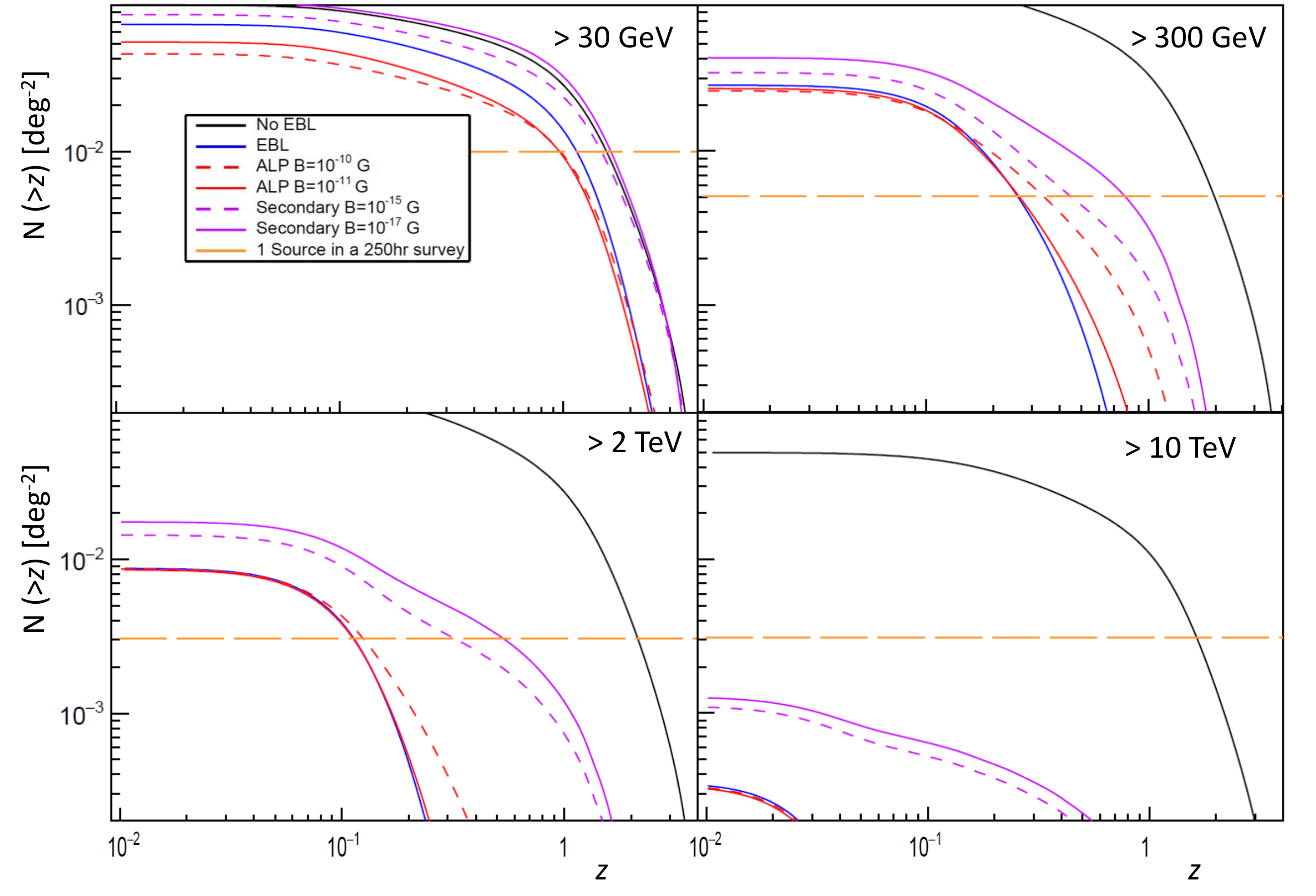} 
\caption{Cumulative source count as a function of redshift for CTA-South, assuming a 5$\sigma$ integral flux sensitivity and 50 hr exposure observation. Dashed horizontal line represents 1 source detected in a 250 hr survey.}
\label{figCumulativeCount_z_50}
\end{figure*}

\begin{figure*}[h!]
  \centering
\includegraphics[width=0.95\textwidth]{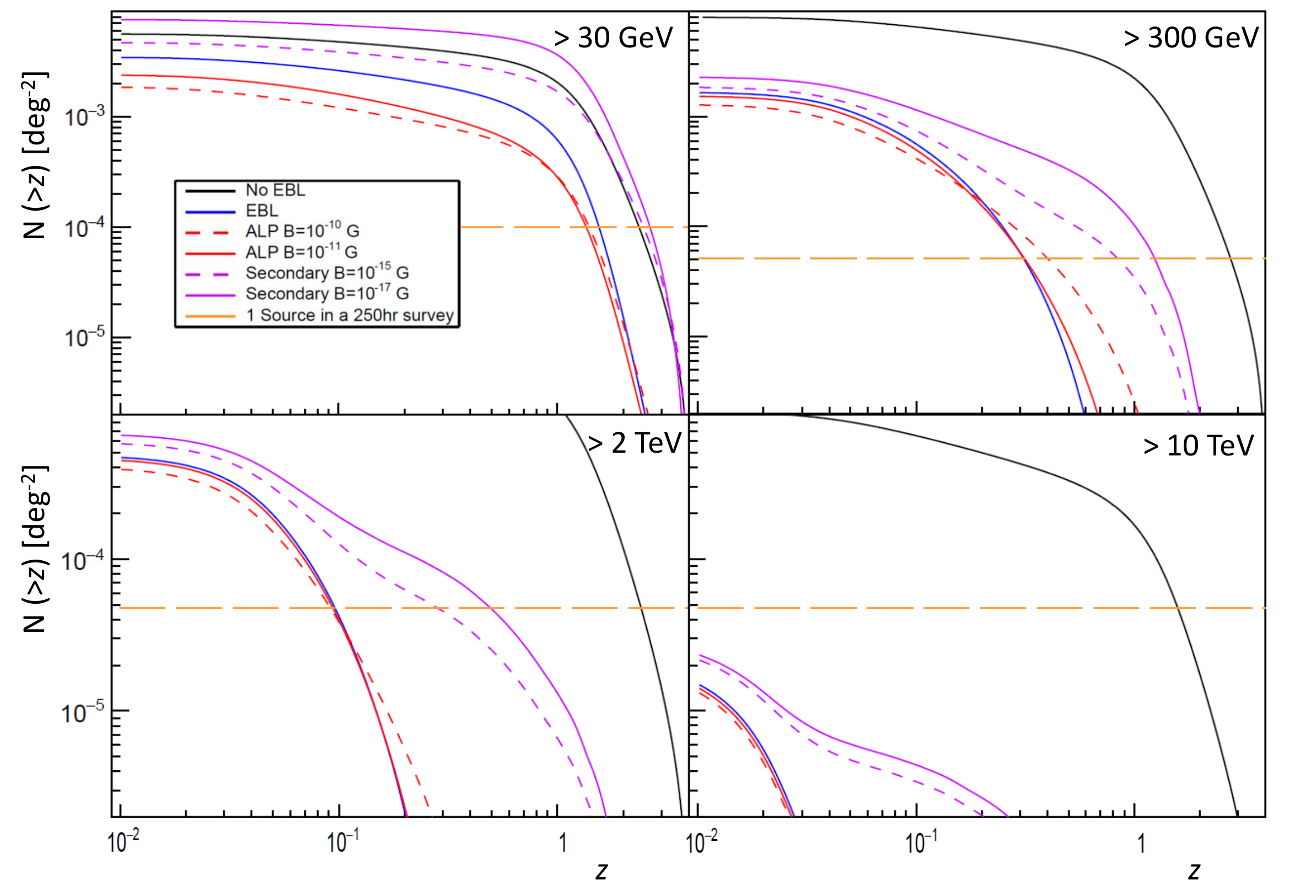} 
\caption{Same as Figure~\ref{figCumulativeCount_z_50}, but assuming 0.5 hr exposure observations.}
\label{figCumulativeCount_z_05}
\end{figure*}

\section{Results}\label{sec:Results}
In Fig.~\ref{figCumulativeCount} we present the blazar cumulative source count distributions for all the above mentioned scenarios in different energy bands. We report the expected flux sensitivity limits for the reference CTA-North and CTA-South (5$\sigma$, 50hr) as derived from~\cite{CTools} with the CTA instrument response function provided by \cite{CTASens}. We also present expected sensitivities of HAWC (5$\sigma$, 1yr)~\cite{HAWCSens}, LHAASO (5$\sigma$, 1yr)~\cite{LHASSOSens} and Hybrid (5$\sigma$, 1yr)~\cite{Hybrid}. 

To optimize the sensitivity over a very broad dynamic range of gamma-ray energies, CTA will comprise telescopes of three different sizes. Respectively from the largest to the smallest they are tuned for the energy range: 20 GeV $\lesssim$ E $\lesssim$ 200 GeV, 100 GeV $\lesssim$ E $\lesssim$ 10 TeV, 1 TeV $\lesssim$ E $\lesssim$ 300 TeV. They are also designed with different fields of view (roughly 5$^{\circ}$, 7$^{\circ}$, 9$^{\circ}$ from the largest to the smallest telescopes. Observation strategies will adopt a combination of sub-arrays with different contributions to the sensitivity of the overall array. The pronounced difference of the sensitivity between CTA-South and CTA-North at high energies is due to the lack of the smallest telescopes in the latter, and at low energies to the presence of fewer middle sized telescopes and a larger geomagnetic field intensity at the CTA-North site~\cite{Szanecki2013,Maier2015}.

In Table~\ref{tabSource}, we summarize the expected source detection counts in different energy bands for a CTA unbiased survey of 250 hours from either the Southern or the Northern location, assuming the standard model with only gamma--ray absorption on the EBL. We consider the cases of 50, 5 and 0.5 hours observation time. For CTA-South, we consider FoV of 5$^{\circ}$, 7$^{\circ}$, 9$^{\circ}$ respectively for observation in the energy bands $>30$ GeV, $>300$ GeV, $>2$ TeV. CTA-North will not host any small sized telescope, thus we assume a FoV of 7$^{\circ}$ even at $>2$ TeV. The portion of the sky covered in a survey of 250 hr with observation of 5 hr each is 2.4\%,4.7\% and 7.7\% respectively for the assumed FoV of 5$^{\circ}$, 7$^{\circ}$, 9$^{\circ}$. The total portion of the sky covered scales with the inverse of the observation time per FoV.

\begin{table}[h!]
\begin{center}
 CTA-South/North -- EBL\\
\begin{tabular}{cccc}
\midrule
Energy Band & & $\Delta$T \\ \midrule
  & 50 hr  & 5 hr & 0.5 hr\\
\addlinespace
$>$30 GeV   &  7/4   &  15/9  &  35/13\\
$>$300 GeV    &  5/3  &  13/8  &  32/15\\
$>$2 TeV    &  3/1  &  7/4  &  10$^*$/4  \\
 \bottomrule
\end{tabular}
\end{center}
\caption{Assuming absorption on the EBL only, estimated source count for a blind survey of 250 hours for CTA-South and CTA-North for different energy bands and $\Delta$T observation time per FoV. Note: `*' this source count corresponds to a survey of half of the sky, which is completed in $\approx$130 hours. }
\label{tabSource}
\end{table}
Interestingly, Table~\ref{tabSource} shows that a shallow and broad survey is preferred to maximize the number of sources identified.
In Fig.~\ref{figCumulativeCount_z_50} and~\ref{figCumulativeCount_z_05} we depict the cumulative source count in redshift for all the models considered in this paper. These figures show results for blazars with an observed flux higher than the CTA-South integral flux sensitivity for 50 hr and 0.5 hr of exposure time per FoV.
From these figures it can be seen, once again, that a survey with shorter observation time per FoV would provide the highest discovery potential. We note that the detection of blazars at $>10$ TeV is hardly expected in the assumed survey modes.

\begin{table*}[h!]
\begin{center}
CTA-South/North -- 0.5 hr\\
\begin{tabular}{ccccc}
\midrule
& $>$30 GeV & $>$300 GeV  & $>$2 TeV   & $>$10 TeV \\
\midrule
\addlinespace
No EBL   &  56/22  &  156/68  &  96$^*$/36  &  20$^*$/8\\
EBL   &  35/13  &  32/15  &  10$^*$/4  &  -/-\\
ALP (B=$10^{-10}$ G)   &  19/7  &  25/12  &  8$^*$/4  &  -/-\\
ALP (B=$10^{-11}$ G)    &  24/9  &  30/14  &  10$^*$/4  &  -/-\\
Secondary (B=$10^{-15}$ G)    &  47/18  &  36/16  &  12$^*$/5  &  1$^*$/-\\
Secondary (B=$10^{-17}$ G)   &  76/32  &  45/20  &  14$^*$/6  &  1$^*$/-\\
\bottomrule
\end{tabular}
\end{center}
\caption{Estimated source count for a blind survey of 250 hours with observations of 0.5 hr with CTA-South and CTA-North in different energy bands and for the models considered in this paper. See text for further details. Note: `-' mark denotes no expected detection; `*' this source count corresponds to a survey of half of the sky, which is completed in $\approx$130 hours}
\label{tabCTA_Model}
\end{table*}

\subsection{On the impact of ALP and secondary gamma--rays}  Secondary gamma rays produce a higher cumulative source count as shown in Fig.~\ref{figCumulativeCount} and in Table~\ref{tabCTA_Model}, while the contribution of ALP seems more modest. Being both effects strongly dependent on the yet unconstrained IGMF strength, no firm conclusion could be drawn on ALPs or secondary photons of hadronic origin from a future CTA data analysis like the one outlined in this paper alone. Nonetheless, Fig.~\ref{figCumulativeCount_z_50} and Fig.~\ref{figCumulativeCount_z_05} suggest that the different gamma-ray propagation scenarios would lead to quite distinctive source distributions in redshift at the assumed CTA flux sensitivity, with possible discovery of sources at $z>1$ at energies up to $>2$ TeV for the secondary gamma-ray scenario, where a simple model including only EBL absorption would fail.

\subsection{Extensive Air Shower observatories} Although HAWC, LHAASO and Hybrid are not as sensitive as CTA in the sub-TeV energy range, they possess much higher FoV and duty cycles, which make them very competitive for blazar discovery at multi TeV. For this reason, we also derived expected source counts for a 5-year survey with these experiments, with the assumption that the integral flux sensitivity simply scales with the square root of the observation time. The results are presented in Table~\ref{tabEAS_Model}. 

Installation of the 300 water Cherenkov tanks of HAWC was completed in 2015. Since then the observatory has been observing at full capacity, providing already first results~\cite{HAWCResults1,HAWCResults2,HAWCResults3,HAWCResults4}, that suggest a possible upgrade scenario with the construction of an additional observatory in the Southern hemisphere~\cite{HAWCSouth}. HAWC has a very wide coverage of the sky (15\% instantaneous, 2/3 over 24 hr) and it has a great potential for survey discovery. We estimated the number of extragalactic sources that HAWC is expected to detect in 5 years of operation. The instrument integral sensitivity is provided in~\cite{HAWCSens} only for gamma rays above 2 TeV, thus we report our results only for this energy range. 

LHAASO is proposed to be composed of different kinds of detectors based on different detection technique. Its flux sensitivity above 10 TeV is dominated by a particle detector array covering an area of $\sim1$~km$^2$\cite{LHASSOSens}. From 300 GeV to 10 TeV, the flux sensitivity is dominated by a combination of water Cherenkov detector array, similar to HAWC, and an array of wide field IACTs. The integral sensitivity used in this work is based on the one reported in~\cite{LHASSOSens} for 1-year observation time with the water Cherenkov detector (sky coverage $\sim$7~steradians) and 50 hr with the IACTs (non steerable). The duty cycle of any IACT is around 1000 hr per year. The FoV of the IACT array of LHAASO is $14^\circ \times 16^\circ$, which means it will take $\sim$5 years for the $\sim$7~steradians portion of sky covered by the companion detector to drift across the IACTs for a total of 50 hr observation per FoV. As it is difficult to disentangle the sensitivity of the two main components of the observatory, here we decided to simply scale the integral sensitivity in~\cite{LHASSOSens} with the square root of the total observational time. 

For the case of Hybrid, we consider a FoV of $4\pi /6$ and we scale the 1-year integral sensitivity reported in~\cite{Hybrid} for 5 years observation in the same way as for the other two above mentioned projects. The target area covered by Hybrid is 10$^4$ m$^2$, and its sensitivity increases roughly proportionally to the square root of the instrumented area. We note that, as shown in Table~\ref{tabEAS_Model}, the effect of ALPs and the secondary gamma-rays is expected to be modest. 

\begin{table*}[h!]
\begin{center}
HAWC / LHAASO / Hybrid -- 5 years\\
\begin{tabular}{ccccc}
\midrule
& $>$300 GeV  & $>$2 TeV   & $>$10 TeV  \\
\midrule
\addlinespace
No EBL   &  -/353/40  & 77/ 685/28  &  -/67/-  \\
EBL   &  -/71/8  &  8/55/3  &  -/1/- \\
ALP (B=$10^{-10}$ G)   &  -/56/6  &  7/48/2  &  -/1/-    \\
ALP (B=$10^{-11}$ G)    &  -/66/8  &  8/53/3  &  -/1/- \\
Secondary (B=$10^{-15}$ G)    &  -/81/9  &  10/76/4  &  -/2/- \\
Secondary (B=$10^{-17}$ G)   &  -/100/11  &  12/90/4  &  -/2/-  \\
 \bottomrule
\end{tabular}
\end{center}
\caption{Estimated source count with 5 years of HAWC/LHAASO/Hybrid data, in different energy bands and for the models considered in this paper. See text for further details. Note: `-' mark denotes a missing result due to integral flux sensitivity not available for the instrument in the specific energy band.}
\label{tabEAS_Model}
\end{table*}

\section{Conclusion}\label{sec:Conclusions}
In this work, we presented the most up to date prediction for the blazar source count distribution for different VHE gamma--ray band between  $>30$ GeV and $>10$ TeV. Table~\ref{tabSource} summarizes the expected number of sources discovered by means of an unbiased extragalactic survey for both CTA-North and CTA-South as a function of the exposure time in the leading scenario of primary gamma rays attenuated on the EBL. Our study confirms that a shallower, more extended survey with observation time of around 0.5 hr per FoV is optimal for CTA in order to maximize the discovery potential of new sources. 

In our work, we also calculated the number of blazars detectable by HAWC, LHAASO and Hybrid (see Table~\ref{tabEAS_Model}) under the assumption of a steady state, averaged SED. Since we use the SED template calibrated by {\it Fermi}, the assumed SED represents the SED averaged over the time. Therefore, we may be able to find weaker objects in flaring state. Although this is applicable to all observatories, Extensive Air Shower observatories have high-duty-cycle, wide FoV, but they require higher fluxes for source detection compared to IACTs, thus the presence of flares would be particularly relevant for this kind of instruments.

We note that the construction site of the two CTA arrays has been recently selected, and a new finely tuned set of Monte Carlo simulations on its performance is to be expected. Previously, the Crab spectrum was used to compute the instrument response; thus, a new study assuming the average SED of~\cite{Ajello2015} would provide more precise results from this work. Furthermore, more realistic predictions can be made by considering the flux sensitivity variations across the FoV of the instrument or different possible divergent pointing modes. The instruments response might differ from the one used in this work, yet the results shown in Fig.~\ref{figCumulativeCount} would still remain valid, with errors arising mainly from statistical uncertainties on the estimation of the parameters of the GLF\footnote{Reported in~\cite{Ajello2015}}. The sole uncertainties on its normalization could lead to a detected source count ranging from 34\% to 230\% compared to the one presented in this work. Also, the predicted number of sources in the multi-TeV might be significantly reduced by the presence of an exponential cut-off in the SED, expected somewhere above few TeV~\cite{Bartoli2012ApJ,Aharonian2003A&A,Krennrich2001ApJ}. Yet, this effect was not included in this work, as the precise energy scale of the cutoff has not yet been uniquely identified. 

We also investigated how the source count distribution would get modified in the presence of ALPs or secondary gamma rays of hadronic origin. Table~\ref{tabCTA_Model} shows examples of the effect that either ALPs or secondary gamma rays may have for the detectable source count for a survey of 0.5 hr per FoV for the two CTA locations. A simple cumulative source count analysis would not be definitive to claim ALPs detection or hadronic acceleration in AGN, yet a multi-wavelength campaign with precise measurement of sources at different redshifts, could provide a hint of any of these two scenarios being at work. However, the case of secondary gamma rays yields a very distinctive source count redshift distribution, with expected detections of sources at $z >$1 at energies up to $>$2 TeV (Figs.~\ref{figCumulativeCount_z_50} and~\ref{figCumulativeCount_z_05}).

In this paper, we focused only on blazars which are the dominant source class in the extragalactic VHE sky. Current IACT observatories also reported discovery of a handful of starburst and radio galaxies (misaligned blazars). It is still difficult to quantitatively estimate detectability of these populations by a survey. Although gamma-ray luminosity functions for starburst galaxies and radio galaxies are available (e.g.~\cite{AckermannAjello2012,InoueRadioGalaxies}), they are still highly uncertain because the sample size is small and they converted luminosity function at different wavelengths using luminosity correlations. Moreover, their SEDs at the VHE band are not well established yet. TeV gamma rays from galaxies are known to be heavily attenuated by internal radiation fields (e.g.~\cite{InoueStarburst}). Centaurus A, a radio galaxy, is found to have a second gamma--ray component at $>$ 5 GeV in addition to a lower gamma--ray energy component~\cite{Sahakyan2013,Brown2016}. The emission mechanism for this second component is still under debate. However, CTA is expected to increase the number of these sources as well with its improved sensitivity~\cite{Acero2013b, Sol2013}, and will potentially lead to the serendipitous discovery of new source populations in the VHE sky~\cite{CRAcc1,CRAcc2,CRAcc3,CRAcc4}.

\section*{Acknowledgement }
ADF is supported by the People Programme (Marie Curie Actions) of the European Union Seventh Framework Programme FP7/2007-2013/ under REA grant agreement n [317446] INFIERI “INtelligent Fast Interconnected and Efficient Devices for Frontier Exploitation in Research and Industry“. YI is supported by the JAXA international top young fellowship and JSPS KAKENHI Grant Number 
\break JP16K13813. MASC is a Wenner-Gren Fellow and acknowledges the support of the Wenner-Gren Foundations to develop his research. This research received support from the Daiwa Foundation Small Grants. GC acknowledges funding from STFC Consolidated Grant ST/N000919/1 and STFC Project Grant ST/M00757X/1. This research made use of ctools, a community-developed analysis package for Imaging Air Cherenkov Telescope data. ctools is based on GammaLib, a community-developed toolbox for the high-level analysis of astronomical gamma-ray data. Data for the secondary gamma rays fluxes were kindly provided by Oleg E. Kalashev. 
The authors wish to thank Alex Kusenko, Masaki Mori, Fabrizio Tavecchio, Manuel Meyer, and Kazunori Kohri for discussions and suggestions; Abelardo Moralejo, Daniel Mazin and Masahiro Teshima for their valuable comments on understanding the performance difference of the two CTA arrays at low energies; Michele Doro for comments and bringing to our attention the novel hybrid detector. This paper has gone through internal review by the CTA Consortium.

\section*{References}
\bibliography{mybibfile_truncated}

\end{document}